\documentclass[11pt]{whitepaper}
\usepackage{times}
\usepackage{graphicx}
\usepackage{latexsym}
\usepackage{fancybox}
\usepackage{color}
\renewcommand{\thepage}{Page~\arabic{page} of 10}
\newcommand{\kader}[2]{
\begin{center}
\fbox{
\parbox{#1}{\begin{center} 
\vspace*{-0.25cm}
#2
\vspace*{-.25cm}
\end{center}}
}
\end{center}}

\evensidemargin\oddsidemargin
\newdimen\oldparindent\oldparindent=0.80em
\newdimen\refindent

\begin{document}

\color{black}

\begin{center}
{\LARGE {\it Kepler\/} White paper:\\[1cm]} {\bf\Huge
\color{blue}E\color{black}nsemble \color{blue}AS\color{black}terose\color{blue}I\color{black}smo\color{blue}L\color{black}ogy of the
  \color{blue}Y\color{black}oung Open Cluster NGC\,2244\\[0.25cm]
  \color{blue}(EASILY)\color{black}\\[0.25cm]} 

\color{black} {\Large Conny Aerts (conny@ster.kuleuven.be), Konstanze Zwintz,
  Pablo Marcos-Arenal, Ehsan Moravveji, Pieter Degroote, P\'eter P\'apics,
  Andrew Tkachenko \& Joris De Ridder, {\it University of
    Leuven, Belgium}\\[0.2cm]

Maryline Briquet \& Anne Thoul, {\it University of Li\`ege, Belgium}\\[0.2cm]

Sophie Saesen, Nami Mowlavi, Fabio Barblan, 
{\it Geneva Observatory, Switzerland}\\[0.2cm]

Coralie Neiner and the MiMeS collaboration, {\it Paris-Meudon Observatory,
  France}\\[0.2cm]

Kresimir Pavlovski, {\it University of Zagreb, Croatia}\\[0.2cm]

Joyce Guzik, {\it Los Alamos National Laboratory, USA} \\[1cm]Abstract\\ }
\normalsize \end{center}

Our goal is to perform in-depth ensemble asteroseismology of the young (age less
than 6 million years) open cluster NGC\,2244 with the 2-wheel {\it Kepler\/}
mission.  While the nominal {\it Kepler\/} mission already implied a revolution
in stellar physics for solar-type stars and red giants, it was not possible to
perform asteroseismic studies of massive OB stars because such targets were
carefully avoided in the FoV in order not to disturb the exoplanet hunting.  Now
is an excellent time to fill this hole in mission capacity and to focus on the
metal factories of the Universe, for which stellar evolution theory is least
adequate.

Our white paper aims to remedy major shortcomings in the theory of stellar
structure and evolution of the most massive stars by focusing on a large
ensemble of stars in a carefully selected young open cluster. Cluster
asteroseismology of very young stars such as those of NGC\,2244 has the major
advantage that all cluster stars have similar age, distance and initial chemical
composition, implying drastic restrictions for the stellar modeling compared to
asteroseismology of single isolated stars with very different ages and
metallicities.

Our study requires long-term photometric measurements of stars with visual
magnitude ranging from 6.5 to 15 in a large FoV with a precision better than
$\sim$30\,ppm for the brightest cluster members (magnitude below 9) up to
$\sim$500\,ppm for the fainter ones, which is well achievable with the 2-Wheel
{\it Kepler\/} mission, in combination with high-precision high-resolution
spectroscopy and spectro-polarimetry of the brightest pulsating cluster
members. These ground-based spectroscopic data will be assembled with the HERMES
and CORALIE spectrographs attached to the twin 1.2\,m Mercator and Euler
telescopes at the Observatories of La Palma, Canary Islands and La Silla, Chile,
respectively, as well as with the spectro-polarimetric NARVAL instrument
attached to the 2\,m Bernard Lyot Telescope at the Pic du Midi in the French
Pyrenees, to which the scientists in the present consortium have guaranteed
access.\\[1cm]

\begin{figure}[h!]
\begin{center}
\rotatebox{0}{\resizebox{12cm}{!}{\includegraphics{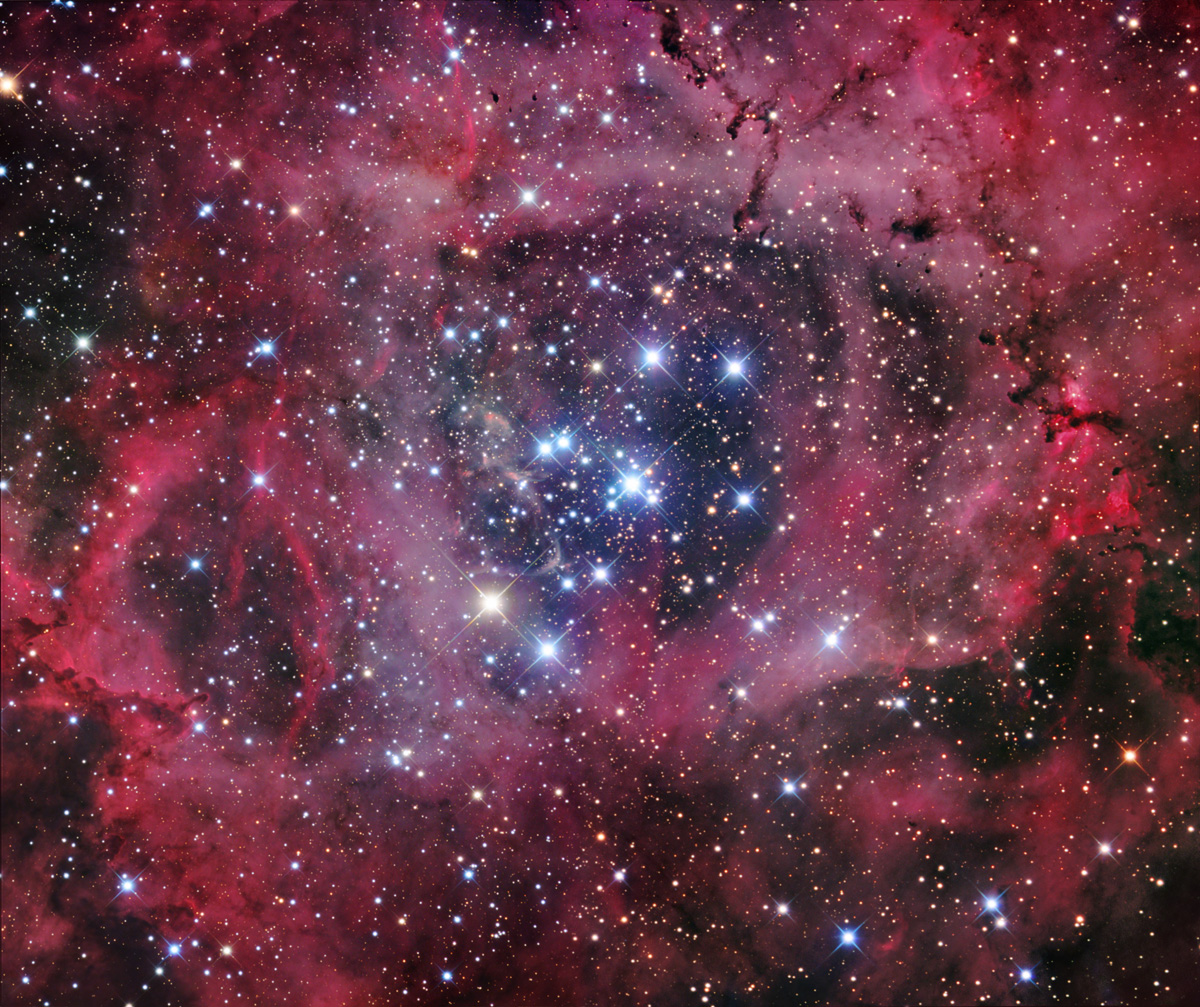}}}
\end{center}
\caption{Image of the open cluster NGC\,2244 on the sky.}
\label{fig1}
\end{figure}


\parindent=0.5cm

\section*{Scientific Context and State-of-the-Art}

Stellar evolution theory is most uncertain for massive stars, where core
convective overshooting, internal differential rotation, magnetic fields, and
transport of angular momentum and of chemical species are all important yet
poorly understood physical ingredients. Moreover, these phenomena are of
relevance during many of the evolutionary phases, from the star formation until
the very late stages.  While parametrised descriptions have been used to include
them in stellar evolution codes of single and binary stars, the theoretical
predictions remain essentially uncalibrated by high-precision data so far. Quite
crucially, the introduction of these physical processes has a major impact on
the outcome of the theory.  Moreover, the predictions for the evolutionary
properties of massive OB stars, and by implication also the theoretical
predictions for supernova progenitors, galaxies, and chemical enrichment of the
Universe as a whole, are considerably different for codes that have been
developed by independent teams. We cannot but conclude that theoretical stellar
evolution models of massive stars are very uncertain while, at the same time,
these models are used as the building blocks of much of modern astrophysics.
Here, we focus on the earliest stages of stellar evolution of massive stars.

$$\star$$$$\star\ \ \ \ \ \star$$

Asteroseismology of the four open clusters in the field-of-view of the nominal
{\it Kepler\/} mission, based on solar-like oscillations in red giants is
successful and progressing fast (Stello et al.\ 2011, Miglio et al.\ 2012,
Corsaro et al.\ 2012). These studies have the potential to lead to seismic
constraints on mass loss on the red giant branch for clusters of different
metallicity and age covering a few to several giga years.  {\it CoRoT nor {\it
  Kepler\/} were able to gather suitable data for ensemble modeling of young
open clusters of only a few million years old, based on OB-type pulsators. This
is the purpose of our proposal.}

Pulsating OB stars in or past the core-hydrogen burning stage reveal low-order
pressure and gravity modes as well as high-order gravity modes excited by
various mechanisms, making it necessary to observe them with a time base of
months in order to resolve a sufficient number of modes needed to perform
seismic modeling.  Large ground-based multitechnique campaigns dedicated to
pulsating B stars had been performed prior to the CoRoT launch in 2006 and have
led to the derivation of the core overshoot parameter and of the ratio of the
core to envelope rotation for a few targets (Aerts et al.\ 2003, 2011; Briquet
et al.\ 2007; Dziembowski \& Pamyatnykh 2008, among others).  Despite large
efforts, daily aliases remained an observational difficulty, because these stars
often have (beat) periods of the order of days and the modelling had to be
restricted to only a few identified modes.

Crucial steps ahead were achieved in the assembly and interpretation of seismic
data of various B and Be stars with a time base of five months and a sampling of
32 seconds with the CoRoT mission. In addition to the derivation of the core
overshoot parameter for a few more stars, this revealed a diversity in stellar
variability of massive stars with amplitude below 0.1\,mmag that was
unanticipated from ground-based data (Neiner et al.\ 2009, Huat et al.\ 2009,
Guti\'errez-Soto et al.\ 2009; Diago et al.\ 2009; Degroote et al.\ 2009, 2010a,
2012; P\'apics et al.\ 2011, 2012).  The duty cycle of the data solved the
aliasing problems occurring in ground-based data and these recent studies led to
compatibility checks for current stellar models (Aerts et al.\ 2011), to the
derivation of the extent of core overshoot and of extra near-core inhomogeneous
near-core mixing (Degroote et al.\ 2010b) and quite often to the failure to
explain the excitation of detected modes from present-day pulsation theory. It
also pointed out that the pulsations of Be stars are in need of better
theoretical interpretation which can handle the complexity of the frequency
spectrum due to the deformation and fast rotation of those stars (Neiner et
al.\ 2012).

{\it To bring a new dimension in this research, we need observations with a high
  duty cycle and with a time base of at least half a year for an ensemble of
  massive stars with similar constraints on metallicity and age, in such a way
  that we can detect rotational splitting of the pulsation modes of numerous
  cluster stars, including the slowest rotators. }

The limits of what can be done from the ground were explored by Saesen et
al.\ (2010), who organised a huge ground-based 2.5-year multicolour photometric
multisite campaign focusing on the young open cluster NGC\,884, involving 12
observatories.  This study revealed 36~multi-periodic B-stars and 39
mono-periodic B-stars, but, despite immense efforts, suffered severly from daily
aliasing and low duty cycles, as well as being limited to light curves with a
precision of 5.7~mmag in the $V$ filter, 6.9~mmag in $B$, 5.0~mmag in $I$ and
5.3~mmag in $U$ for the brightest cluster members.  From frequency analysis and
empirical mode identification from amplitude ratios of five cluster members,
Saesen et al.\ (2013) managed to deduce a seismic age for the cluster in
agreement with the one deduced from an eclipsing binary cluster member. Their
study was a proof-of-concept for the power of asteroseismic modeling of a young
open cluster, should data of higher duty cycle and with a factor ten better
precision become available. At the same time, {\it this study made it obvious
  that progress cannot be made efficiently from ground-based efforts because we
  need better detection capabilities for the pulsation modes, typically at the
  level of 0.1 to 0.5\,mmag.}

A similar conclusion holds for young stars in their pre-main-sequence (pre-MS)
phase: progress only started efficiently by monitoring members of young open
clusters with the MOST and CoRoT satellites, in combination with ground-based
spectroscopy and in some cases X-ray measurements (e.g., Zwintz et al.\ 2011,
2013a,b). Variability at the level of 0.5\,mmag or less occurs in a diversity of
pre-MS stars while theoretical instability computations fail to explain the
detected modes for various stars.

\kader{6.2in}{\bf We conclude that high-precision ensemble seismic modeling of
  the most massive stars and of stellar objects near the birthline is
  unavailable at present. With this project, we will provide it with the 2-Wheel
  {\it Kepler\/} mission for a carefully selected young open cluster containing
  confirmed pulsators in the appropriate classes among its members.}

\begin{figure}[t!]
\begin{center}
\rotatebox{0}{\resizebox{12cm}{!}{\includegraphics{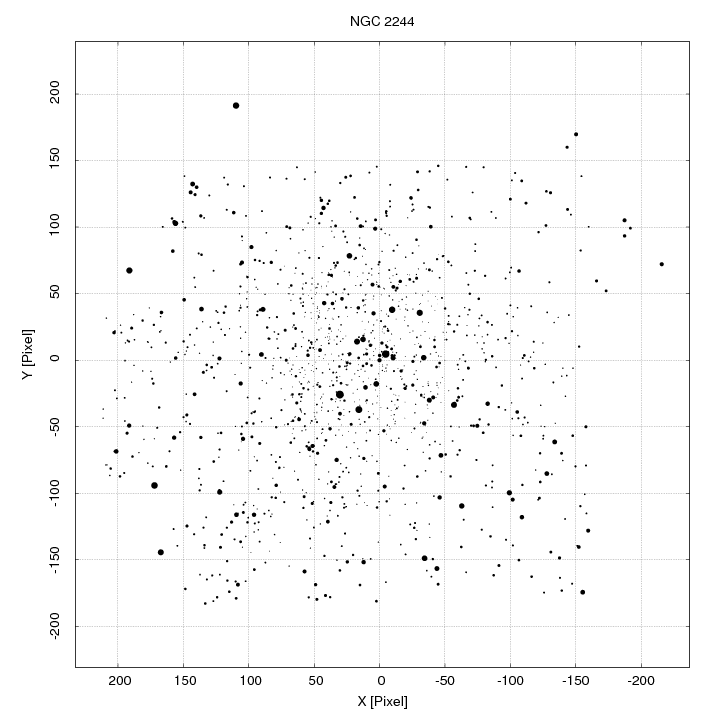}}}
\end{center}
\caption{Cluster map obtained from the WEBDA data base of open clusters. The
  scale is such that 6.00 units correspond with one arcminute (source: Massey et
  al.\ 1995).}
\label{fig2}
\end{figure}

\begin{figure}[t!]
\begin{center}
\rotatebox{0}{\resizebox{12cm}{!}{\includegraphics{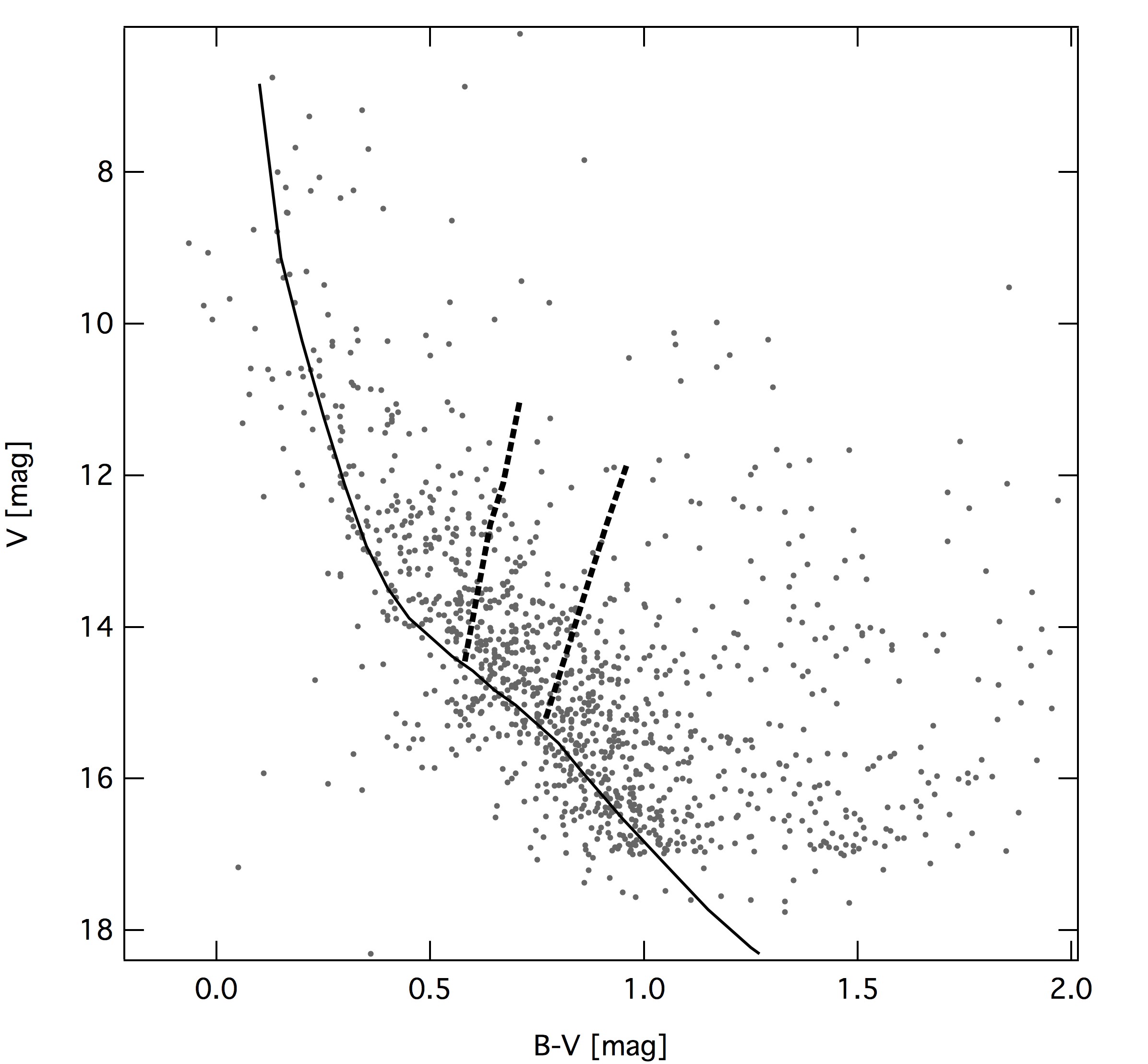}}}
\end{center}
\caption{Colour-magnitude diagram of all the known members of the open cluster
  NGC\,2244. The full line indicates the zero-age-main-sequence while the two
  dotted lines represent the blue and red border of the classical instability
  strip.}
\label{fig3}
\end{figure}

\section*{The main target: the young open cluster NGC\,2244}

The very young open cluster NGC\,2244 (Fig.\,1; RA2000 = 06:31:55, DEC2000 =
04:56:30) belongs to the Mon OB2 association and is situated at a distance of
1.6$\pm$0.2\,kpc. Its age is estimated to be between one and six million years
(Bonatto \& Bica 2009).  Its diameter is some 30 arcminutes and its stellar
density is suitable for observations with {\it Kepler\/} in the sense that
crowding is not a major limitation (Fig.\,2). Several members of this bright
cluster have also been studied carefully in UV-wavelengths and in X-rays, which
will allow us to get the basic stellar properties from well-determined spectral
energy distributions. Such multiwavelength studies revealed a discrepancy
between spectroscopic and evolutionary masses of cluster members with masses
below 25\,M$_\odot$ (Martins et al.\ 2012).

Chemical abundance analyses of hot massive cluster members (Vranken et
al.\ 1997; Pavloski \& Hensberge 2005; Martins et al.\ 2012) agree very well
with the present-day standard abundances of $X=0.715$ and $Z=0.014$ determined
by Przybilla et al.\ (2008). Hence, pulsations are expected all along the main
sequence of NGC\,2244 (Miglio et al.\ 2007). We shall pay specific attention to
the occurrence or absence of pulsations in stars with spectral type near B9 to
A0, where new pulsators were only recently found from ground-based cluster
studies (Mowlavi et al.\ 2013), while such periodic variability is not predicted
from theoretical models.

A major reason why we selected NGC\,2244 as the best target for our scientific
aims is the fact that four of its O-type stars have already been monitored by
CoRoT during a short run lasting 34 days. This revealed a diversity of causes
for the detected stellar variability:
\begin{itemize}
\item
The slowly-rotating O9V member HD\,46202 (visual magnitude of 8.2) turned out to
be a ``classical'' $\beta\,$Cep-type pulsator, the one with the highest mass
known so far (Briquet et al.\ 2011). Its pulsation frequencies occur near
60$\mu$Hz and have amplitudes between 30 and 100\,ppm.  While
its few detected zonal modes could be well modeled by current stellar models,
none of the observed modes are predicted to be excited and a discrepancy was
found between the spectroscopic and seismic $\log\,g$. Rotational splitting was
not detected in the time series, due to a too short time base resulting in too
poor frequency resolution.

\item
Quite unexpectedly, stochastic oscillations with frequencies between 30 and
80$\mu$Hz and amplitudes between 20 and 50\,ppm with Lorentzian profils similar
to those found in solar-type stars, were found in the primary of the bright
spectroscopic cluster binary HD\,46149 (visual magnitude of 7.6, Degroote et
al.\ 2010a). Just like HD\,46202, this massive pulsator is also a slow
rotator. Its stohastic pulsation modes seem to point out that this star must
somehow sustain a considerable outer convective layer, which might create
magnetic activity close to the stellar surface (Cantiello \& Braithwaite
2011). Further investigation is needed to interpret the observed variability;
rotational splitting was not detected so far.

\item
The most massive cluster members HD\,46223 (visual magnitude of 7.2) and
HD\,46150 (visual magnitude of 6.8) are moderate to fast rotators and turned out
to be variable with an excess power at low frequencies and amplitude near mmag,
connected with a yet unknown stochastic process (Blomme et al.\ 2011). It is
noteworthy that similar behaviour was recently identified in the ultra-bright
{\it Kepler\/} GO eclipsing binary B1III target V380\,Cygni (Tkachenko et
al.\ 2012).
\end{itemize}
In addition, gravity-mode pulsations were found in two additional relatively
bright cluster members with spectral type mid-B (visual magnitudes near 10 and
11) from a 21-day MOST campaign (Gruber et al.\ 2012). Modeling was not possible
due to lack of frequency resolution and mode identification, as it concerned
very few isolated frequency peaks, without any signature of rotational
splitting.

Ensemble seismic modeling of these massive cluster members, imposing a single
initial composition and age as in Saesen et al.\ (2013)  was hitherto
impossible due to too limited frequency resolution and lack of mode
identification. However, these preliminary findings show that NGC\,2244 is
ideally suited as target for our scientific goals.

All these previously gathered space photometric data are available to the
proposers of this white paper and {\it will allow us to calibrate the new data
  on NGC\,2244 to be assembled with 2-Wheel {\it Kepler\/}}, with a guarantee to
detect time-resolved pulsations in various of the cluster members.

\kader{6.2in}{\bf We aim to observe the $\sim$700 cluster members brighter than
  visual magnitude 15.  This sample includes numerous pre-MS stars of spectral
  types B and A as well as all the core-hydrogen burning stars of type O and
  B. This will allow us to get the very first complete picture of all the
  variable stars considerably more massive than the Sun in an ultra-young
  cluster and study evolutionary and pulsational effects from the star formation
  process up to the exhaustion of core-hydrogen burning for single and binary
  cluster members. Long-term high-precision photometry revealing the variability
  of young and active stellar objects still embedded partly in their
  circumstellar birth cloud will be addressed as well and confronted with
  existing low-resolution data at infra red wavelengths.  A search
  for hot Jupiters and other planets around A-type stars in this young cluster
  with solar metallicity will be undertaken as well.}

\section*{Observing strategy}

The cluster members have a visual magnitude range between 6.5 and 17 and cover
OBAF stars in the core-hydrogen burning phase as well as lots of pre-MS stars
(with visual magnitude as of 12 or fainter). The colour-magnitude diagram shown
in Fig.\,3 shows the breath in types of stars in the cluster. The twenty-ish
brightest OB stars need to be observed with a precision of some tens of ppm.
They will lead to saturated pixels, but experience with the design and
observations through customized masks for such bright targets in the nominal
{\it Kepler\/} mission has proven to be very successful (Metcalfe et al.\ 2012,
Tkachenko et al.\ 2012). We thus propose to devise specific and very large
elongated masks for those bright cluster members which are brighter than the
saturation limit, compensating also for the drift and jitter motion. Adapted
masks should thus involve thousands of pixels per bright cluster star, but this
is still very modest compared to the more than five million available pixels.

From the {\it Call for white papers document\/} we deduce that 2-Wheel {\it
  Kepler\/} can point to an accuracy of about one arcsec, undergoes a drift of
120 arcsec/day and maintains a one arcsec jitter about the drift line for less
than or equal to a maximum of 4 days, after which a repointing is
necessary. From Fig.\,2, we derived the minimum requirement to limit the
acceptable drift to 1 arcminute, i.e., we propose to repoint every 12 hours. In
this way, we avoid that the photons of various cluster members get mixed into
the masks of neighbouring stars while on the other hand we do not interfere with
the periods of the pressure modes (typically between 3 and 7 hours). The gravity
modes have periods typically between one and three days, which is too long to
monitor completely without repointing. Half a day seems to be the best
compromise to leave the cluster stars in their masks while having suffiently
long time strings to perform {\it a posteriori\/} corrections for inter- and
intrapixel variability. For the latter, we shall rely on the methods for jitter
(Drummond et al.\ 2006) and background (Drummond et al.\ 2007) correction
algorithms developed previously in the Leuven team for the CoRoT mission. This
software shall be adapted to the case of the 2-Wheel {\it Kepler\/} operations.

{\it An alternative operational mode would be to consider very many small
  adjacent masks.  Each target would then go out of a particular mask and go
  straight into the next mask, etc. Appropriate flux book-keeping might allow to
  achieve one merged time series per target, which would avoid having lots of
  stars in one large elongated mask. It is to be investigated with the Kepler
  instrument team whether this is an acceptable mode of operation; if so then
  repointing at a different pace than requested above could be considered.}

Previous experience with asteroseismology of B pulsators from ground-based
multisite campaigns and from CoRoT data (e.g., Degroote et al.\ 2012) has shown
that rotational splitting in slowly rotating OB pulsators as some stars in
NGC\,2244 requires a frequency resolution of at least 0.06$\mu$Hz.  {\bf This is
  an essential requirement, because successful seismic modeling demands
  unambiguous identification of the pulsation modes and high-precision frequency
  matching. A shorter time base will only lead to the discovery and
  classification of the variables in the cluster but will not allow seismic
  modeling capable of improving the interior physics of the stars, while this is
  our global aim.  Thus, we request continuous monitoring of the field centered
  on NGC\,2244 during six months.}

Should our white paper be accepted, then we shall organise simultaneous
ground-based time-resolved spectroscopic and spectro-polarimetric campaigns with
the twin 1.2m Mercator \& Euler telescopes, and with the 2m Bernard Lyot
telescope. The proposing team has permanent access to these three facilities.
This will be appropriate for the brightest cluster members only (visual
magnitude up to 9). For the fainter cluster members, we shall apply for time on
larger aperture telescopes (polarimetric mode of the 3.6m telescope (HARPSpol)
and VLT (high-resolution multifiber UVES/FLAMES spectrograph) available at ESO
through the normal biannual competition for which the current team members were
very successful the past years.

Priority should be given to devise the optimal masks for all the $\sim$700
NGC\,2244 cluster members brighter than magnitude 15, which can be observed with
a precision better than $\sim$500\,ppm.  Using all 5.44 million pixels takes
about 12 minutes of time to readout and store on board. Even though such slow
readout is an acceptable cadence for our science case, it is better to adopt a
higher read-out cadence for less pixels, in order to optimally resolve the
pulsational behaviour of all the cluster members.  Thus, we adopt the strategy
to read out all the pixels of all the cluster members with a cadence of 5
minutes, keeping many of the pixels available for simultaneous complementary
science, several of which can be read out in short cadence.

\section*{Concrete Workplan and Timeline}

Let us denote the start of the 6-months monitoring with 2-Wheel {\it Kepler\/}
observations by $T_0$. The timeline of our project is as follows:
\begin{enumerate}
\item
$[T_0,T_0$+6 months$]$: gathering of ground-based spectroscopy and
  spectro-polarimetry. Computation of suitable grids of evolutionary models with
  the MESA code (Paxton et al.\ 2011) and with the accompanying non-adiabatic
  pulsation code GYRE (Townsend \& Teitler 2013). 
\item
$[T_0$+6 months,$T_0$+12 months$]$: calibration and analysis of {\it Kepler\/}
  photometry and ground-based data. Classification of all the variables in
  NGC\,2244 and selection of all the pulsators. First publication on the data
  and variables.
\item
$[T_0$+12 months,$T_0$+18 months$]$: binary light curve and radial-velocity
  modeling; identification of the wavenumbers $(\ell,m)$ of all the detected
  pulsation frequencies. Second publication on the pulsators and their modes.
\item
$[T_0$+18 months,$T_0$+24 months$]$: ensemble modeling of the entire
  cluster. Publication of the final modeling results.

\end{enumerate}
All the necessary methodology and software for these tasks is available at the
host institutes of the proposers. We refer to the Reference list below as a
witness of the experience of the current proposers in this exciting research
field.

\section*{Complementary Science Cases}

Our proposal is focused on NGC\,2244 as {\bf the main target}. However, the {\it
  Kepler\/} FoV is about 100 square degrees and lots of additional
scientifically valid targets in or near the Galactic disc will come for free
within this FoV, while centering it on the main cluster target. Should our white
paper be selected for observations, we suggest to make a call to the community
to propose additional targets and operation modes in terms of left-over masks
and integration times, with the coordinate information of NGC\,2244.

It is noteworthy that the FoV can be chosen such as to
have overlap with one of the short-run pointings done by CoRoT and that a
classification of the variability of faint CoRoT stars was produced by members
of our current team; all those CoRoT data are available to us (Debosscher et
al.\ 2011). In general, the Leuven classification software is available for any
of the 2-wheel {\it Kepler\/} pointings to optimize the scientific output of the
mission.

In this way, the present white paper will become a community proposal with the
prime science case of NGC\,2244 and numerous additional science cases
complementary to the one we propose here. Surely, the {\it Kepler\/} community
will \color{blue} EASILY \color{black} come up with brilliant ideas for the free
pixels in the surrounding area near NGC\,2244.

\section*{References}
{\small
C.\ Aerts et al., {\it Science}, {\bf 300}, 1926 (2003)\\
C.\ Aerts et al., {\it Astron. Astrophys.\/}, {\bf 534}, id.A98 (2011)\\
R.\ Blomme, et al., {\it Astron. Astrophys.\/}, {\bf 533}, A4 (2011)\\
C.\ Bonatto, E.\ Bica, {\it Mon. Not. of the Roy. Astron. Soc.\/}, {\bf 394},
2127 (2009)\\
M.\ Briquet et al., {\it Astron. Astrophys.\/}, {\bf 466}, 269 (2007)\\
M.\ Briquet et al., {\it Astron. Astrophys.\/}, {\bf 527}, A112 (2011)\\
M.\ Cantiello, J.\ Braithwaite, {\it Astron. Astrophys.\/}, {\bf 534}, A140
(2011)\\
E.\ Corsaro et al., {\it The Astrophys. J.\/}, {\bf 757}, id.190 (2012)\\
J.\ Debosscher, et al.,  {\it Astron. Astrophys.\/}, {\bf 529}, A89 (2011)\\
P.\ Degroote, et al., {\it Astron. Astrophys.\/}, {\bf 506}, 111 (2009)\\
P.\ Degroote, et al., {\it Astron. Astrophys.\/}, {\bf 519}, A38 (2010a)\\
P.\ Degroote, et al., {\it Nature\/}, {\bf 464}, 7286, 259 (2010b)\\
P.\ Degroote, et al., {\it Astron. Astrophys.\/}, {\bf 542}, A88 (2012)\\
P.\,D.\ Diago, et al.,  {\it Astron. Astrophys.\/}, {\bf 506}, 125 (2009)\\
R.\ Drummond, et al., {\it Pub. of the Astron. Soc. of the Pacific\/}, {\bf
  118}, 844 (2006)\\
R.\ Drummond, et al., {\it Astron. Astrophys.\/}, {\bf 487}, 1209 (2008)\\
W.\,A., Dziembowski, A.\,A.\ Pamyatnykh, {\it  Mon. Not. of the
  Roy. Astron. Soc.\/}, {\bf 385}, 2061 (2008)\\
D.\ Gruber, et al.,  {\it Mon. Not. of the Roy. Astron. Soc.\/}, {\bf 420}, 219
(2012)\\
J.\ Guti\'errez-Soto, et al.,  {\it Astron. Astrophys.\/}, {\bf 506}, 133 
(2009)\\
A.-L.\ Huat, et al., {\it Astron. Astrophys.\/}, {\bf 506}, 95 (2009)\\
F.\ Martins, et al., {\it Astron. Astrophys.\/}, {\bf 538}, A39 (2012)\\
D.\ Massey, et al.,  {\it The Astrophys. J.\/}, {\bf 454}, 151 (1995)\\
T.\,S.\ Metcalfe, et al., {{\it The Astrophys. J. Letters\/}, 
{\bf 748}, id.L10 (2012)\\
A.\ Miglio, et al., {\it Mon. Not. of the Roy. Astron. Soc.\/}, {\bf 375}, L21
(2007)\\
A.\ Miglio, et al., {\it Mon. Not. of the Roy. Astron. Soc.\/}, {\bf 419}, 2077
(2012)\\
N.\ Mowlavi, et al., {\it Astron. Astrophys.\/}, {\bf 554}, A108 (2013)\\
C.\ Neiner, et al., {\it Astron. Astrophys.\/}, {\bf 506}, 143 (2009)\\
C.\ Neiner, et al., {\it Astron. Astrophys.\/}, {\bf 546}, A47 (2012)\\
P.\,I.\ P\'apics, et al.,  {\it Astron. Astrophys.\/}, {\bf 528}, A123 (2011)\\
P.\,I.\ P\'apics, et al.,  {\it Astron. Astrophys.\/}, {\bf 542}, A55 (2012)\\
B.\ Paxton, et al., {\it The Astrophys. J. Suppl. Series}, {\bf 192}, id.3
(2011)\\ 
K.\ Pavlovski, H.\ Hensberge, {\it Astron. Astrophys.\/}, {\bf 439}, 309 
(2005)\\
N.\ Przybilla, et al., {\it The Astrophys. J.\/}, {\bf 688}, L103 (2008)\\
S.\ Saesen, et al., {\it Astron. Astrophys.\/}, {\bf 515}, 16 (2010)\\
S.\ Saesen, et al., {\it Astron. J.\/}, in press (2013, arXiv:1307.4256)\\ 
D.\ Stello, et al., {\it The Astrophys. J.\/}, {\bf 739}, id.13 (2011)\\
A.\ Tkachenko, et al.,  {\it Astron. Astrophys.\/}, {\bf 424}, L21 (2012)\\
R.\,H.\,D.\ Townsend, S.\ Teitler,  {\it Mon. Not. of the Roy. Astron. Soc.\/},
in press (2013, arXiv:1308.2965)\\
M.\ Vranken, et al., {\it Astron. Astrophys.\/}, {\bf 320}, 878 (1997)\\
K.\ Zwintz, et al.,  {\it The Astrophys. J.\/}, {\bf 729}, id.20 (2011)\\
K.\ Zwintz, et al., {\it Astron. Astrophys.\/}, {\bf 550}, A121 (2013a)\\
K.\ Zwintz, et al., {\it Astron. Astrophys.\/}, {\bf 552}, A68 (2013b)\\
}
\end{document}